\documentclass[]{spie}
\pdfoutput=1

\usepackage{amsmath,amsfonts,amssymb}
\usepackage{graphicx}
\usepackage[colorlinks=true, allcolors=blue]{hyperref}

\usepackage{rotating}
\usepackage[export]{adjustbox}

\newcommand{\crossbar}{\texttt{crossbar}}
\newcommand{\crossbario}{\texttt{crossbar.io}}
\newcommand{\grafana}{\texttt{Grafana}}
\newcommand{\influxdb}{\texttt{InfluxDB}}

\newcommand{\twisted}{\texttt{twisted}}
\newcommand{\asyncio}{\texttt{asyncio}}
\newcommand{\docker}{\texttt{docker}}
\newcommand{\ocs}{\texttt{ocs}}
\newcommand{\socs}{\texttt{socs}}
\newcommand{\spt}{\texttt{spt3g\_software}}
\newcommand{\rogue}{\texttt{rogue}}

\title{The Simons Observatory: Overview of data acquisition, control, monitoring, and computer infrastructure}

\author[a]{Brian~J.~Koopman}
\affil[a]{Department of Physics, Yale University, New Haven, CT 06520, USA}

\author[b]{Jack~Lashner}
\affil[b]{University of Southern California, Los Angeles, CA 90007, USA}

\author[a]{Lauren~J.~Saunders}
\author[c]{Matthew~Hasselfield}
\affil[c]{Center for Computational Astrophysics, Flatiron Institute, New York, NY 10010, USA}

\author[d]{Tanay~Bhandarkar}
\affil[d]{Department of Physics and Astronomy, University of Pennsylvania, Philadelphia, PA 19104, USA}

\author[a]{Sanah~Bhimani}

\author[e,f]{Steve~K.~Choi}
\author[f]{Cody~J.~Duell}
\affil[e]{Department of Astronomy, Cornell University, Ithaca, NY 14853, USA}
\affil[f]{Department of Physics, Cornell University, Ithaca, NY 14853, USA}

\author[g]{Nicholas~Galitzki}
\affil[g]{Department of Physics, University of California, San Diego, CA 92093-0424, USA}

\author[h]{Kathleen~Harrington}
\affil[h]{Department of Astronomy and Astrophysics, University of Chicago, 5640 South Ellis Avenue, Chicago, IL 60637, USA}

\author[i]{Adam~D.~Hincks}
\affil[i]{Department of Astronomy \& Astrophysics, University of Toronto, 50 St. George St., Toronto ON M5S 3H4, Canada}

\author[j]{Shuay-Pwu~Patty~Ho}
\affil[j]{Department of Physics, Stanford University, CA 94305, USA}

\author[a]{Laura~Newburgh}

\author[k]{Christian~L.~Reichardt}
\affil[k]{School of Physics, University of Melbourne, Parkville, VIC 3010, Australia}

\author[g]{Joseph~Seibert}

\author[g]{Jacob~Spisak}

\author[l]{Benjamin~Westbrook}
\affil[l]{Department of Physics, University of California, Berkeley, CA 94720, USA}

\author[d,m]{Zhilei~Xu}
\affil[m]{MIT Kavli Institute, Massachusetts Institute of Technology, Cambridge, MA 02139, USA}

\author[d]{Ningfeng~Zhu}

\authorinfo{Corresponding author: \href{mailto:brian.koopman@yale.edu}{brian.koopman@yale.edu}}

\usepackage{aas_macros}

\begin{document} 
\maketitle

\begin{abstract}
The Simons Observatory (SO) is an upcoming polarized cosmic microwave
background (CMB) survey experiment with three small-aperture telescopes and one
large-aperture telescope that will observe from the Atacama Desert in Chile. In
total, SO will field over 60,000 transition-edge sensor (TES) bolometers in six
spectral bands centered between 27 and 280 GHz to achieve the sensitivity
necessary to measure or constrain numerous cosmological parameters, including
the tensor-to-scalar ratio, effective number of relativistic species, and sum
of the neutrino masses. The SO scientific goals require coordination and
control of the hardware distributed among the four telescopes on site. To meet
this need, we have designed and built an open-sourced platform for distributed
system management, called the Observatory Control System (\ocs{}). This control
system interfaces with all subsystems including the telescope control units,
the microwave multiplexing readout electronics, and the cryogenic thermometry.
We have also developed a system for live monitoring of housekeeping data and
alerting, both of which are critical for remote observation. We take advantage
of existing open source projects, such as \crossbario{} for RPC and PubSub,
\twisted{} for asynchronous events, \grafana{} for online remote monitoring,
and \docker{} for containerization. We provide an overview of the SO software
and computer infrastructure, including the integration of SO-developed code
with open source resources and lessons learned while testing at SO labs
developing hardware systems as we prepare for deployment.
\end{abstract}


\section{INTRODUCTION}
\label{sec:intro}

The Simons Observatory (SO) is a new cosmic microwave background (CMB)
experiment being constructed in the Atacama Desert in northern Chile. The
observatory will consist of three small aperture telescopes (SATs) and one
large aperture telescope (LAT). Spread among these telescopes will be over
60,000 cryogenic bolometers, measuring the temperature and polarization of the
CMB from 27 to 280 GHz in six frequency bands.  These detectors will be read
out using a microwave multiplexing ($\mu\mathrm{MUX}$)
architecture.\cite{hendersonHighlymultiplexedMicrowaveSQUID2018} The
combination of small and large aperture telescopes was chosen to cover a wide
range of science goals, which include measuring primordial tensor
perturbations, further characterizing primordial scalar perturbations,
constraining the sum of the neutrino masses and effective number of
relativistic species, measuring the thermal and kinematic Sunyaev-Zel'dovich
effects for a large sample of galaxy clusters, and measuring the duration of
reionization.\cite{the_simons_observatory_collaboration_simons_2018,
2019BAAS...51g.147L}

Orchestrating the control, data collection, and monitoring across all
telescopes, their associated readout electronics, and the ancillary
housekeeping (HK) electronics is a critical task for any observatory. Past CMB
experiments have developed their own control codes or used those of similar
experiments, for instance the General Control Program (GCP) developed initially
on the Sunyaev-Zel'dovich Array and used across several other experiments
including SPT, the Keck-array, BICEP2, and PolarBear. \cite{Story_2012,
2014ApJ...792...62B} The Atacama Cosmology Telescope (ACT) developed its own
remote control and monitoring infrastructure, which evolved over the years
since the initial deployment in 2007 \cite{10.1117/12.790006,
swetz_overview_2011, thornton_atacama_2016, Koopman:2018vgj}. Generally, these
control softwares have been closed source, and insufficient for the scale of
hardware deployed on SO.

For the Simons Observatory we have designed and written a distributed control
system we call the Observatory Control System (\ocs{}). \ocs{} is a modular system,
designed to make it easy for any user to add additional hardware to the system.
This design structure enables straightforward adaptation of existing code where
similar hardware is used, and to extend the list of supported hardware for use
on existing and new deployments of \ocs{}. 

In these proceedings we present an overview of \ocs{} in Section
\ref{sec:overview}, describing its components, how it is deployed, and its
development process. Then, in Sections \ref{sec:agents} and \ref{sec:clients} we
describe ``Agents'' and ``Clients'', the core components of \ocs{}, and
describe various \ocs{} ``Agents'' and ``Clients'' that are written or planned.
Next, in Section \ref{sec:live_monitoring} we describe the live monitoring
capabilities of \ocs{}. In Section \ref{sec:deployment} we describe three
different deployments; namely, a simple deployment, a deployment in one of the SO
collaboration labs, and the planned deployment at the site in Chile, before
concluding in Section \ref{sec:summary}. Appendix \ref{sec:acronyms} contains
Table \ref{tab:acronyms} which lists all acronyms used throughout the text for
reference.

\section{OCS OVERVIEW}
\label{sec:overview}

The \ocs{}\footnote{\url{https://github.com/simonsobs/ocs}} is a distributed
control system designed to coordinate data acquisition in astronomical
observatories. A design goal has been ease of use in managing hardware
operations and I/O tasks in distributed systems like those found at an
observatory, where multiple processes must come together to record an
observation. For a system to be easy to use, it should have a shallow learning
curve, allow for users to add or modify components for their use cases, and
have a simple installation procedure. The size of an \ocs{} installation can be
small, such as a single computer in a test lab, or complex such as a
multi-computer system at the observatory.  Throughout the rest of the paper, we
refer to a single computer or CPU as a ``node.''

\subsection{Architecture}

\begin{figure} [ht]
\begin{center}
\begin{tabular}{c}
\includegraphics[width=0.8\linewidth]{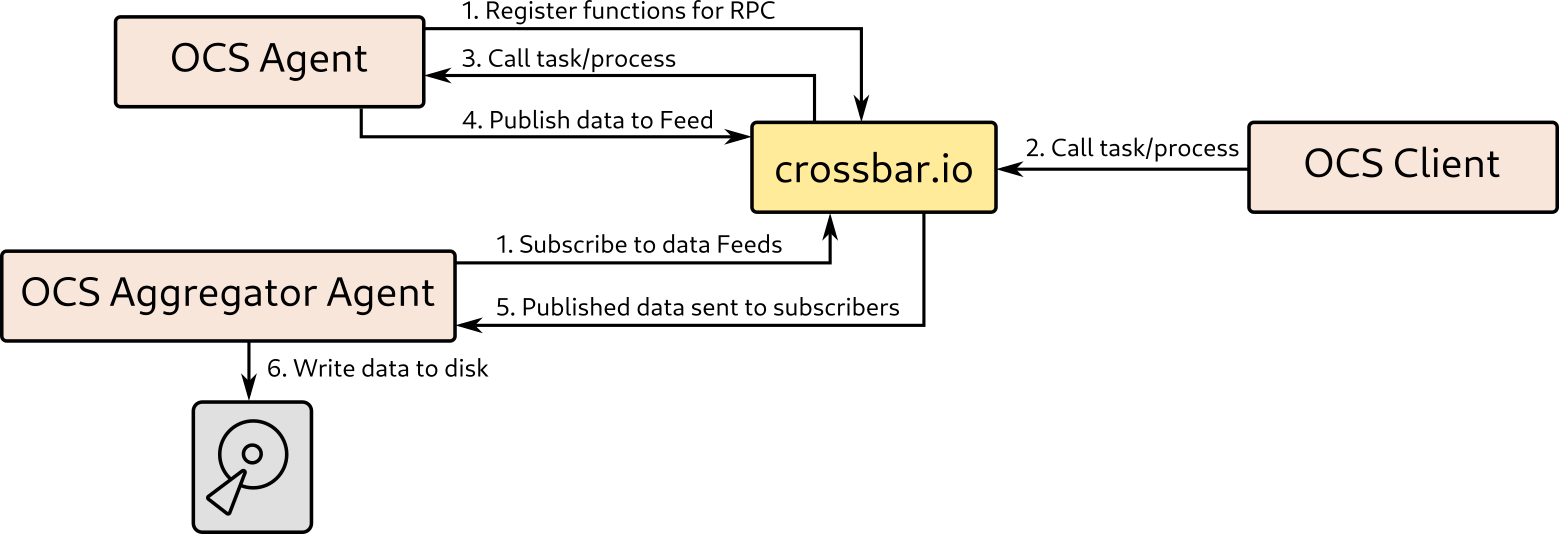}
\end{tabular}
\end{center}
\caption[example]
{ \label{fig:ocs_interactions}
The interaction of two \ocs{} Agents (a generic Agent, and the HK Aggregator Agent
that writes data to disk), an \ocs{} Client, and the \crossbar{} server. Upon
startup (1) the Agents both register their tasks and processes for remote
calls and the \ocs{} Aggregator Agent subscribes to all data Feeds, (2) an \ocs{}
Client makes a call to an agent task or process, (3) the crossbar server
forwards this call to the Agent, (4) the Agent publishes data to an \ocs{} Feed,
(5) the HK Aggregator, a subscribed Agent, receives a copy of the data, and (6)
writes the data to disk.}
\end{figure}

The \ocs{} has two primary components: \ocs{} Agents and \ocs{} Clients. \ocs{}
Agents are long-running software servers which interface with a hardware or
software component to coordinate data acquisition. Each Agent contains a set of
operations, referred to as ``Tasks'' and ``Processes'' that run for predefined
or open-ended amounts of time, respectively. Tasks and Processes may be
configured to run at Agent startup or may be initiated by a remote procedure call
(RPC) over the network by an \ocs{} Client. Information is passed among
Agents on the network via a Publish/Subscribe (PubSub) pattern. Both RPC and
PubSub are performed through a central Web Application Messaging Protocol
(WAMP) router. The specific WAMP router implementation we have chosen is
\crossbario{}\footnote{\url{https://crossbar.io}} (referred to in this paper as
just \crossbar{}). A diagram showing the interaction between Agents, Clients,
and the crossbar server is shown in Figure \ref{fig:ocs_interactions}.

\ocs{} Clients are scripts that orchestrate the actions of one or more \ocs{}
Agents on the network. Clients can take many forms, as \crossbar{} supports
several different programming languages. Most commonly, they are Python scripts
run by the user on the command-line or JavaScript routines running in a web
browser. Clients can also be a part of an Agent, commanding other Agents when
needed. A good example of this is the Observation Sequencer described in
Section \ref{sec:sequencer}.

Clients can be run anywhere on the network that has access to the \crossbar{}
server, such as a dedicated computer at the site or a user's laptop. Agents can
be run directly on bare metal; however more commonly they will be run within
Docker containers for ease of setup, use, and monitoring, as described in
Section \ref{sec:docker}.

\subsection{Dependencies}

We leverage several open source tools in \ocs{}, which we describe in this
section.

\crossbar{} is an open source implementation of the Web Application Messaging
Protocol (WAMP) used for distributed applications.
WAMP\footnote{\url{https://wamp-proto.org/}} is a websockets subprotocol that
provides two application messaging patterns, RPC and PubSub. Developed by
Crossbar.io (the company), \crossbario{} (the software) is accompanied
by the Autobahn libraries which provide open source WAMP implementations in
several programming languages. This was appealing when selecting a mechanism
for RPC and PubSub, as it allowed support for Python, JavaScript, and, if
needed, C++. These Autobahn libraries provide the base components which allow
one to write WAMP ``application components'' that we use to run the \ocs{}
Agent code, described in more detail in Section \ref{sec:agents}.

The WAMP application components are necessarily asynchronous. Autobahn offers
the choice of either of the \asyncio{} or \twisted{} libraries for implementing
the asynchronous networking framework; we use the latter.

We selected the SPT3G software (\spt{})
framework\footnote{\url{https://github.com/CMB-S4/spt3g_software}} for the \ocs{}
data storage format \cite{harrington_thesis}. The code is written in C++
with a light Python layer. The internal structure of the files consist of
``Frames'', which are processed through ``Pipelines'' by ``Modules'' one at a
time. We have built a new library,
\texttt{so3g},\footnote{\url{https://github.com/simonsobs/so3g}} which operates
on these standard G3 frames, taking advantage of the structure of the
underlying \spt{} file format, while providing helper methods for loading data
without users needing to consider the frame structure.

Housekeeping (HK) data are collected and written to SPT3G files (or
\texttt{.g3} files) by the HK Aggregator Agent, where frame length and file
length are configurable parameters. Detector data are collected by a G3
pipeline called the ``SMuRF Streamer" and are sent over the network to the SMuRF
Recorder Agent, where they are written to \texttt{.g3} file on disk. For more
information on these Agents see Section \ref{sec:agents}.

\subsection{Configuration}
The Agents and connection information for the WAMP router are defined within an
\ocs{} Site Configuration File (SCF). This is a simple YAML file, containing the
address for the WAMP router, some internal WAMP address information, and a list
of each Agent, grouped by which host they will run on. Each Agent is defined
by its ``agent-class'', a unique name given to each type of Agent and an
``instance-id'', a unique name given to a particular instance of a type of Agent
(the latter is necessary because many copies of a single Agent may be running
on the network, say for interfacing with multiple identical pieces of
hardware). Command-line arguments to further configure Agents are passed to
the Agent as a list.

The SCF, together with the Docker Compose configuration files that define
the Docker containers in which individual \ocs{} Agents run (see Sec.
\ref{sec:docker}), completely define the \ocs{} network. This makes the system
incredibly easy to migrate to a different machine, or to restore to a machine
that might need to be replaced due to a hardware failure or any other reason.
In addition, an expert is often able to identify misconfiguration by viewing
just these two configuration files.

If an Agent is deployed outside of the Docker framework, the SCF defines
the paths to the Agent's code. These paths may be split into different
repositories. The \ocs{} repository provides core Agents critical to the
functionality of \ocs{}, while Agents specific to the Simons Observatory
hardware are kept in a separate repository called \socs{} (the Simons Observatory
Control System).\footnote{\url{https://github.com/simonsobs/socs}}

\subsection{Docker}
\label{sec:docker}

We use Docker extensively in the deployment of \ocs{} within the Simons
Observatory. It allows for the packaging of software, libraries, and
configuration files for each Agent into shareable Docker images. These
isolated, reproducible environments for each Agent have been valuable in the
deployment of \ocs{} in labs throughout the SO collaboration. Agents are built
into images and published to Docker
Hub,\footnote{\url{https://hub.docker.com/u/simonsobs}} an online resource for
hosting and downloading public Docker images. This is done automatically by the
Continuous Integration pipelines described in Section \ref{sec:CI}.

Docker Compose is used to start up the multi-container environment needed for
\ocs{}. Long-running services such as the \texttt{crossbar} server,
\influxdb{}, and \grafana{} are managed separately, and configured to start at
boot. This allows for the \ocs{} stack to be restarted in a straightforward
manner without affecting these third party services. Orchestration tools such
as Docker Swarm, for managing docker containers across the network are still
being explored, though we note that Swarm does not support device connections
like those needed for USB hardware such as the Lakeshore 240 (a thermometry
readout system).

\subsection{Open Source, Documentation, and User Contributions}

Clear documentation is vital for adoption of any software package. Since the
core of \ocs{} is written in Python we make use of the Sphinx
tool\footnote{\url{https://www.sphinx-doc.org/en/master/}} for creating our
documentation, which is freely hosted on Read the
Docs,\footnote{\url{https://readthedocs.org/}} an automatic building,
versioning, and hosting platform for open source project documentation. Each
commit to the \ocs{} repositories triggers a new build of the documentation
which
is automatically updated on the Read the Docs website, and publicly viewable.

Documentation is split into several sections, including a User guide that
provides a walk-through of the setup of an \ocs{} network, the Agent reference
detailing what each \ocs{} Agent does and how to use and configure them, and a
Developer guide describing in more detail how core components of \ocs{} work
and how users can develop their own Agents and Clients. This thorough
documentation has been key to the development and use of \ocs{}, including among the
test labs within the Simons Observatory.

The core of \ocs{} and the first Agents were initially developed by a small team
of developers. This allowed for rapid development of the structure of \ocs{}
and demonstration of the ability to interface with specialized hardware in the
lab. The first example use case was temperature readout and control of a
$^{3}\mathrm{He}$/$^{4}\mathrm{He}$ dilution refrigerator, primarily through
the associated low noise resistance bridges and cryogenic temperature modules
produced by Lakeshore Cryotronics.\footnote[1]{\url{https://www.lakeshore.com}}

\setcounter{footnote}{1}

As test labs deployed more and more hardware during the design and
building phase of SO they began using \ocs{} installations to facilitate
hardware testing. It is much easier to develop and debug new Agents when one
has direct access to the requisite hardware. Thus, researchers in multiple
labs developed new Agents for their specific hardware by modeling their work
on existing Agents with the help of the \ocs{} documentation. This practice has
made the integration of hardware into the experiment relatively
straightforward. It has also been valuable for developing \ocs{} since we have
had several testing environments to work out bugs in the core \ocs{} code.

\subsection{CI Pipeline}
\label{sec:CI}

The development of \ocs{} has benefited from the use of Continuous Integration
(CI), namely, the practice of frequently integrating code to a shared
repository, triggering the automated building and testing of code changes. We
initially used Travis CI as our CI platform, but recently migrated to GitHub
Actions, which integrates nicely with GitHub and allows simpler control over
workflows triggering on specific events. The initial motivation was to automate
the building of updated Docker images for the Agents upon new commits.  Before
the use of continuous integration, these images would need to be built by hand
and pushed to a Docker registry for distribution, an easy step to forget during
development.

There are several workflows defined with the \ocs{} repositories. The first
focuses on testing. \ocs{} has a set of unit tests written to be run with
pytest.  The workflow will build the Docker images for each \ocs{} component
and run the unit tests within an \ocs{} container. Code coverage is reported,
extracted from the container, and published to the Coveralls
platform.\footnote{\url{https://coveralls.io/}} Finally the documentation is
built within a container to test for any failures in that process, as we have
found it easy to break Sphinx builds on Read the Docs, especially when a new
python module dependency is added. This workflow is valuable as it catches any
bugs introduced by new changes that might break the unit tests. It is run on
any commit and pull request, providing rapid feedback to contributing
developers.

The other workflows focus on the building and pushing of Docker images to
Docker Hub. We distinguish between the stable releases of \ocs{} and the
development branch of \ocs{}. Development images are built and pushed on merge
commits to the ``develop'' branch. These have descriptive tags and allow for
deployment of images under testing before the next official versioned release.
Official releases of \ocs{} trigger a separate workflow which also builds Docker
images with succinct tags corresponding to the release version, as well as the
common ``latest'' tag for Docker images. In both workflows these images are
pushed to Docker Hub. Stable images are only built on the main \ocs{} branch,
which is branch-protected and only merged to on scheduled releases.

\section{OCS Agents} \label{sec:agents}

The \ocs{} Agent is one of two major components of \ocs{}. Each \ocs{} Agent is
a subclass of the Autobahn ApplicationSession class. Connection and
disconnection events that occur during the application life cycle
are overridden to provide \ocs{} functionality. On startup, these handle the
registration of tasks and processes for remote calls from \ocs{} Clients, the
startup of any configured task or process, and the startup of the Agent
heartbeat, a signal of Agent health. On shutdown they handle the stopping of
any running tasks or processes, and shut down of the Agent. If an Agent for some
reason becomes disconnected from the \crossbar{} server, it attempts to
reconnect for a short period of time before shutting down altogether. This
allows for brief outages of the \crossbar{} server, say, for instance, if it
needs to be restarted (though in practice the need for this is rare).

The \texttt{OCSAgent} class provides useful wrappers for PubSub and for
providing the Agent operation commands.  Each agent that will publish data must
first register the communication channel, which we call the \ocs{} Feed, a
class which handles the WAMP address space and the buffering and publication of
data submitted to it by an Agent. Once a Feed is registered, the Agent can
publish data to it in ``blocks", each of which contains co-sampled data,
whether for a single timestamp or a longer duration of time. The block
structure is not allowed to change after construction. Its structure, published
to a Feed, is a simple Python dictionary. These data are serialized before
being sent over the network. Prior to publication, the Feed performs some
verification of the message data structure. This check at the Feed level can be
useful for debugging when developing an Agent.

Any agent, or supporting client, can subscribe to an \ocs{} Feed.  Agents that
aggregate all data from other Agents, like the HK Aggregator or InfluxDB
Publisher Agent, subscribe to all \ocs{} Feeds that are marked for recording to
disk.

Tasks and Processes made available for remote call in an Agent have a simple
operation control API that provides four possible actions: ``start", ``status",
``wait", and ``abort" or ``stop" (for Tasks and Processes, respectively).
``Start'' will request that a task or process be run, passing any parameters
needed for the operation. ``Wait'' will block until the operation has exited or
a timeout has elapsed. ``Status'' returns the current status of the operation.
Possible values are ``starting'', ``running'', ``stopping'', and ``done''. The
returned object can also include informative data, such as the most recent data
collected by the Agent in a structure we call the operation session data. The
``abort'' or ``stop'' methods will abort the task or process. These operation
control actions are called by \ocs{} Clients.

In the remainder of this Section we present details on several of the major
\ocs{} Agents, including those used for pointing the telescope, commanding the
readout electronics, writing data to disk, and orchestrating observations. We
also present a summary of all other supporting Agents.

\subsection{Antenna Control Unit}

The Antenna Control Unit (ACU) is an industrial PC used to control the motion
of a single telescope platform. The Simons Observatory will use one ACU per
telescope, provided by Vertex
Antennentechnik,\footnote{\url{https://www.vertexant.com/}} the company that
has been commissioned to build the telescope platforms. Vertex Antennentechnik
additionally provides software on the ACUs that provides an API for controlling
the telescope platform, allowing users to switch between the operational modes
required for observations. The ACU also publishes two data streams to ports on
a computer provided by SO: a 200\,Hz UDP stream containing time stamps and
encoder values, and a 5\,Hz TCP stream containing those same values in addition
to monitors for errors and faults.

The ACU \ocs{} Agent publishes the data received from each of these streams to an
\ocs{} Feed for recording to file and for viewing in the live monitor. The Agent
also provides Operations for commanding the telescope. The functioning of the
incoming data monitors has been verified during factory acceptance testing for
the first commissioned SAT at the Vertex Antennentechnik facility in Germany;
testing of some tasks remains ongoing.

\subsection{SMuRF Agents} \label{ssec:smurf_agent}

The Simons Observatory will read out its detector array via superconducting
quantum interference device multiplexing ($\mu$MUX) with the SLAC
Microresonator Radio Frequency (SMuRF) warm electronics system.
\cite{kernasovskiySLACMicroresonatorRadio2018,hendersonHighlymultiplexedMicrowaveSQUID2018}.
Communication with the SMuRF system is facilitated by the
\rogue{}\footnote{\url{https://github.com/slaclab/rogue}} C++ software library.
\rogue{} implements low-level communication with the SMuRF FPGA, providing
register access and asynchronous data streaming.  It includes an Experimental
Physics and Industrial Control System (EPICS) server that allows register
reading and writing operations to be executed from high-level clients, such as
the Pysmurf library, a Python library provided by SLAC for interfacing with the
SMuRF systems.\footnote{\url{https://github.com/slaclab/pysmurf}}

A dedicated SMuRF server is used to run \rogue{}, Pysmurf, and related \ocs{} Agents,
and to temporarily store auxiliary data before it is copied to a more permanent
storage node where it is archived along with the \ocs{} housekeeping data.
For detector data acquisition, we are using a custom \rogue{} plugin that collects
detector data and relevant metadata through the asynchronous streaming
interface. Data are grouped into chunks of about one second and packaged
into serializable frames using the SPT3G data format. Frames are sent over the
network to the storage node where they are received by the SMuRF Recorder \ocs{}
Agent that writes them to disk.

Control of the SMuRF system and high-level analysis is managed with the
Pysmurf library, which utilizes the EPICS interface to control the hardware.
In development, Pysmurf is primarily controlled interactively through
an iPython interface or a Jupyter notebook. Once common procedures have been
established, functions can be run through \ocs{} using a Pysmurf Controller agent
that manages the execution of Pysmurf scripts.

Pysmurf generates important companion data such as plots and summaries that
describe detector parameters and operating behavior.
This companion data must also be archived along with the detector data, which
is done through the use of the Pysmurf Monitor and Pysmurf Archiver \ocs{} agents.
Pysmurf passes the auxiliary file metadata to the Pysmurf Monitor
Agent running on the server, which then writes the metadata to a database
on the storage node. The Pysmurf Archiver runs on the storage node and
monitors the database for new metadata entries and then copies the
corresponding files over to be archived along with the detector and
housekeeping data.

\subsection{Observation Sequencer}
\label{sec:sequencer}

The Observation Sequencer Agent orchestrates observatory systems to perform
observations. In this sense it is also an \ocs{} Client, issuing commands to other
Agents on the \ocs{} network. The sequencer requests the survey plan from a
separate program and executes a series of commands, e.g. to the various SMuRF
Agents to ready the detectors, to the ACU Agent to point the telescope.

The Sequencer will record its performed actions in a database to support
downstream monitoring. This will be used to flag errors during data
acquisition, confirm the data for a given observation were acquired, associate
the data on disk with the given survey plan, and group data files and
calibration operations together. The Sequencer is currently in a development
phase.

\subsection{Additional Hardware Agents}
\label{sec:other_agents}

Table \ref{table:agents} shows a summary of other \ocs{} Agents that have been
written or are under development for use on the Simons Observatory. These
Agents are easily shared among other \ocs{} users and several are
user-contributed. Beyond the Agents listed here are more to be developed,
including Agents to interface with Fourier Transform Spectrometers,
polarization grid calibrators, cooling loops, additional power supplies, and
more.

\begin{sidewaystable}[ht]
\caption{\label{table:agents}Summary of current and in development \ocs{} Agents.
Core \ocs{} Agents are marked by an asterisk. All other Agents listed are kept with
\socs{}.}
\label{tab:fonts}
\begin{center}
\begin{tabular}{|l|l|p{0.1\linewidth}|p{0.40\linewidth}|}
\hline
\textbf{Agent} & \textbf{Hardware/Software} & \textbf{Development Status} & \textbf{Description}\\
\hline
AggregatorAgent$^\mathrm{*}$ & Housekeeping Archive & In use & Saves HK data from feeds to \texttt{.g3} files in the HK archive.\\
\hline
BlueforsAgent & Bluefors LD400 Dilution Refrigerator & In use & Parses and passes logs from LD400 software to \ocs{}.\\
\hline
CryomechCPAAgent & Cryomech CPA Model Compressors & In use & Communicates with compressor over Ethernet to record operations statistics. On/Off control still in development.\\
\hline
CHWPAgent & Custom Built Cryogenic HWP & Under development & Command and control cryogenic half wave plate (CHWP) hardware.\\
\hline
DSEAgent & Deep Sea Electronics Controller & Under development & Collect diesel generator statistics over modbus interface.\\
\hline
HostMasterAgent$^\mathrm{*}$ & \ocs{} Agents & In use & Optional Agent to start and stop Agent instances. Useful for running Agents outside of Docker containers.\\
\hline
iBootbarAgent & iBootBar PDUs & In development & Monitor and control managed power distribution units (PDUs) for remote power cycling of electronics.\\
\hline
InfluxDBAgent$^\mathrm{*}$ & Influx Database & In use & Record HK data feeds to Influx database for viewing in Grafana.\\
\hline
Lakeshore372Agent & Lakeshore 372 Resistance Bridge & In use & 100 mK thermometer readout and heater control for dilution refrigerator operation.\\
\hline
Lakeshore240Agent & Lakeshore 240 Input Module & In use & 1K-300K thermometer readout.\\
\hline
LabJackAgent & LabJack T4/T7 & In use & Communicates with a LabJack over Ethernet for generic DAC. Used in warm thermometry readout.\\
\hline
PfiefferAgent & Pfieffer TPG 366 & In use & Six channel pressure gauge controller readout over Ethernet. Used to monitor pressures within vacuum systems of the dilution refrigerators.\\
\hline
MeinbergM1000Agent & Meinberg LANTIME M1000 & In use & Monitor health of Meinberg M1000 timing system via the Simple Network Management Protocol (SNMP).\\
\hline
RegistryAgent$^\mathrm{*}$ & \ocs{} Agents & Under development & Tracks the state of other running \ocs{} Agents on the network.\\
\hline
SCPIPSUAgent & SCPI Compatible PSUs & In use & Command and control power supplies compatible with Standard Commands for Programmable Instruments communication protocol.\\
\hline
SynaccessAgent & Synaccess PDUs & In use & Monitor and control managed PDUs for remote power cycling of electronics.\\
\hline
UPSAgent & UPS Battery Backups & In development & Monitor state of SNMP compatible UPS battery backups.\\
\hline
\end{tabular}
\end{center}
\end{sidewaystable}

\section{OCS Clients}
\label{sec:clients}

\ocs{} Clients are the programs that command single or multi-agent sites,
orchestrating device operation and data collection across the observatory. Most
commonly these will be Python scripts, JavaScript embedded in a users web 
browser, or subroutines within \ocs{} Agents. Clients can connect to the 
\crossbar{} server either via websockets, for instance, using the 
\texttt{wampy} python package, or via HTTP or HTTPS. In order for clients to
support subscription, they must connect via websockets and not the HTTP bridge.

Within \ocs{} there are two methods for writing a Python based Client: the
Basic Client, in which the user must define each operation they wish to call,
and the Matched Client, which performs this definition automatically, matching
operation calls to their defined task or process names. By default Basic
Clients connect via websockets, while MatchedClients connect via HTTP, though
both backends are supported in either case. A Matched Client, communicating
over HTTP is perhaps the simplest Client for getting started quickly, and for
simple interactions with running Agents. Clients can be run from anywhere on
the network where the \crossbar{} server is run, allowing for users to send
commands from their laptops as simply as from on-site computing infrastructure.
Javascript clients are discussed in Sec. \ref{sec:ocsweb}.

\subsection{OCSWeb}
\label{sec:ocsweb}

\begin{figure} [p]
  \centering
  \includegraphics[width=0.9\textwidth,frame]{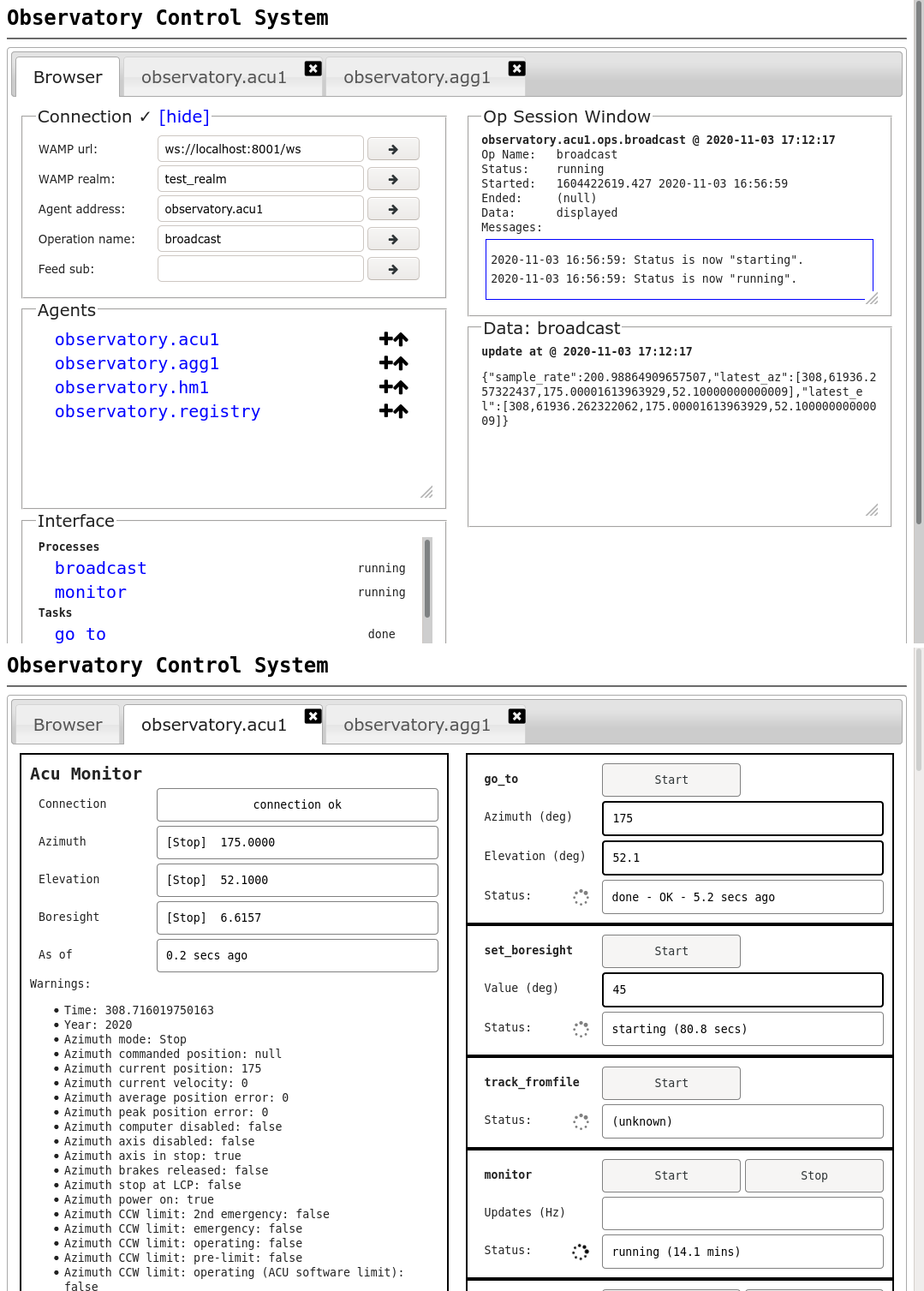}
  \caption{OCSWeb main page (top) and special control panel for the telescope platform control Agent (bottom).}
  \label{fig:ocsweb}
\end{figure}

OCSWeb is an ensemble of \ocs{} Clients implemented in JavaScript and rendered as
a GUI in a web browser. Two screenshots are shown in Figure \ref{fig:ocsweb}.
The elements of OCSWeb are a JavaScript library, HTML and CSS files to present
a basic interface and styling, and a variety of Agent-specific JavaScript
modules to produce tailored interfaces.

For basic functionality, the browser must be able to connect to crossbar through the websocket interface.  The main page of the interface presents a configuration box and a list of active \ocs{} Agents.  Users may interactively select different Agents and browse through the Tasks, Processes and Feeds presented by each one, including the latest session data and log information.  The main page does not permit any detailed control but is useful for basic health monitoring, debugging and development purposes.

Control panels that are tailored to specific Agent types can be launched from the main interface, with each appearing in a new tab of the main window.  The control panels are rendered according to JavaScript code developed specifically for that Agent class.  As is the case for the Agent code itself, the creation of Agent-specific panels is largely left in the hands of developers working with those Agents and the associated hardware.  However, the OCSWeb library provides a framework for generating basic GUI controls and displays, along with examples of how to attach handlers to dispatch Agent requests and process returned information.  The goal is that Agent developers not need to learn much JavaScript to produce a useful control panel.  The basic library is somewhat limited, but does not preclude experienced JavaScript developers from creating more sophisticated, customized control panels should the need arise.

Plans are in development to support persistent configuration between sessions through some combination of a configuration server and HTTP cookies; this will permit views involving certain subsets of Agents to persist from session to session.

\section{Live Monitoring}
\label{sec:live_monitoring}

Remote live monitoring of telescope systems is critical to ensuring high
observation efficiency. Our goal has been to provide a web interface for real
time viewing of housekeeping systems that remote observers can use to assess
the health of the observatory. These systems generally consist of slow data
rate timestreams, on the order of several Hz sampling rates or slower. Higher
data rate timestreams, such as the raw detector data, are more suitable for
custom local interfaces. We searched for existing tools to solve this problem,
and have chosen several open source tools.

\subsection{Grafana}

\begin{figure} [ht]
\begin{center}
\begin{tabular}{c}
\includegraphics[width=0.95\linewidth,frame]{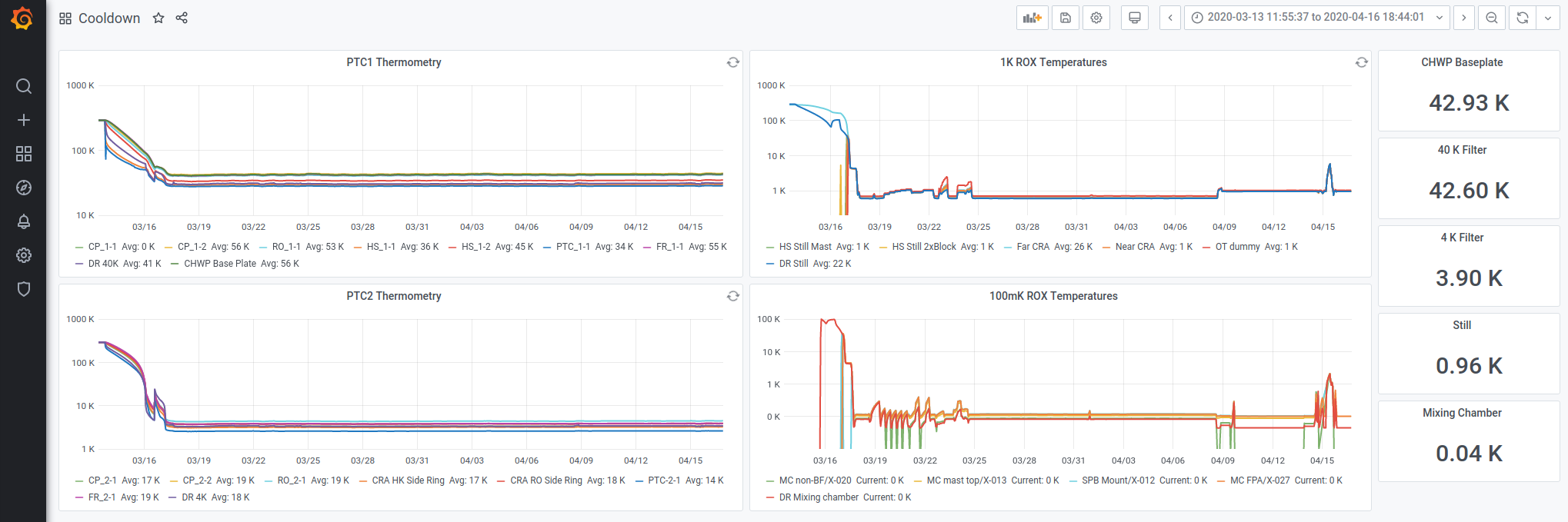}
\end{tabular}
\end{center}
\caption[example]
{\label{fig:grafana-screenshot}
A screenshot of the Grafana live monitor showing cryogenic thermometry
during a cooldown of the SAT1 receiver at UCSD. The y-axis in all plots is
temperature in Kelvin and the x-axis is time, showing approximately a month
of time during which the cryostat was cooled and several tests were run.}
\end{figure}

Grafana\footnote{\url{https://grafana.com/}} is an open source observability
web application used for visualization and analytics of time series data. It
supports many time series database and relational database backends as ``data
sources". Once a data source is configured, and data are inserted into the
database, Grafana can query the database and display the resulting data in a
user's web browser dynamically.  This is done on web pages called
``dashboards", in plots called ``panels". A screenshot showing a Grafana
dashboard with several panels is shown in Figure \ref{fig:grafana-screenshot}.

Dashboards are persistent and can be loaded upon revisiting the web
application. They can be configured to automatically refresh, and provide a
modern web interface for timeseries plots. We have selected InfluxDB as our
primary data source backend. All housekeeping data within \ocs are published
through the crossbar server via \ocs Feeds. There are two primary subscribers to
these feeds: the HK Aggregator and the InfluxDB Publisher. These both record
separate copies of the HK data, to \texttt{.g3} files and to the Influx Database,
respectively.

\subsection{InfluxDB}
\label{sec:influxdb}
InfluxDB, a time series database developed by
InfluxData,\footnote{\url{https://www.influxdata.com/}} is a popular backend
data sources for Grafana, and is designed for fast access to large volumes of
time series data. Hardware running InfluxDB must be sized appropriately,
depending on the number of writes per second and the number of queries per
second. We have found the performance suitable on moderate modern desktop
computers typically found in labs for the slow housekeeping data within SO on
shorter timescales, from one day to several months. Loading a significant
amount of data can cause issues, so one must be aware of the queries configured
within Grafana. We are exploring solutions to this: see Section
\ref{sec:monitoring_challenges}.

\subsection{Log Aggregation with Loki}
\label{sec:loki}
Loki is a log aggregation system designed by Grafana Labs. Written by the same
team that made Grafana, it integrates very well with the Grafana web interface.
A logging driver for Docker publishes logs to the Loki service. Users can then
follow logs or query for past logs directly within the Grafana web interface.
Aggregation of logs external to Docker can be performed using a tool called
\texttt{promtail}, also from Grafana Labs, however most of our requirements are
met by the Docker logging driver so we we are not using it at this time. This
may change in the future.

\begin{figure} [ht]
\begin{center}
\begin{tabular}{c}
\includegraphics[width=0.9\linewidth,frame]{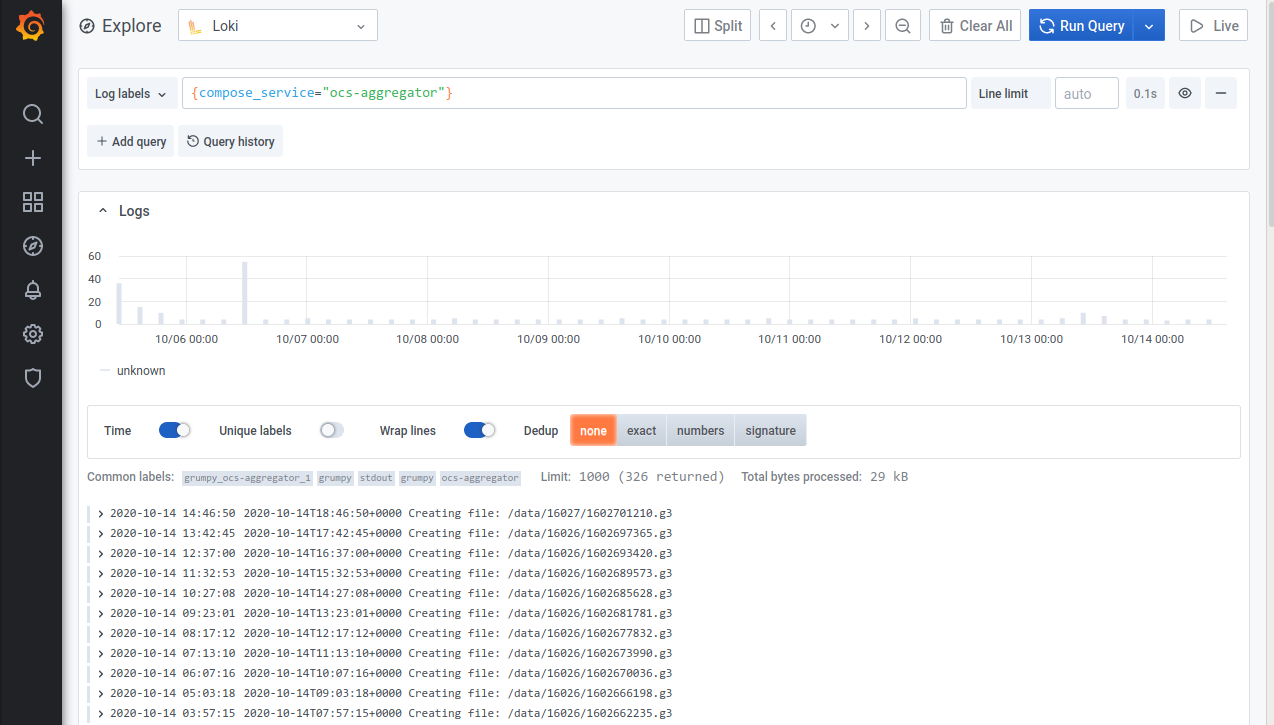}
\end{tabular}
\end{center}
\caption[example]
{\label{fig:loki-screenshot}
A screenshot of the Loki logs within Grafana showing the output of the \ocs HK
Aggregator agent. The top plot shows log frequency over time, and the bottom is
a live feed of the log output. All Agents running within Docker containers can
be configured to output logs to this interface.}
\end{figure}

The interface allows for individual containers, or entire Compose services to
be viewed. \ocs{} logs simply print to \texttt{stdout}, which results in
logging to Docker's configured logging driver. Logs remain visible via the
``docker logs" command, while also being aggregated within Loki. This allows
for easy viewing of logs in the event of an error somewhere within the \ocs
system in a cohesive interface. The interface is shown in Figure
\ref{fig:loki-screenshot}.

\subsection{Detector Data}
The detector data volume is large enough that the HK live monitoring system
cannot support the full detector data rate. The Pysmurf Monitor Agent has the
ability to send a small subset of downsampled detector timestreams to the HK
live monitoring system for basic checks. The full data rate detector
timestreams can be monitored locally via a custom tool from the South Pole
Telescope called
lyrebird.\footnote{\url{https://github.com/SouthPoleTelescope/lyrebird}}
Lyrebird is being adapted for use on SO.

\subsection{Challenges and Future Work}
\label{sec:monitoring_challenges}
Ideally we will retain all HK data in InfluxDB at the full rate throughout the
lifetime of the experiment. However, as mentioned in Sec. \ref{sec:influxdb},
querying InfluxDB for long time ranges can cause issues, even with moderate
data rates. Furthermore, given the limited resolution of any screen,
requesting the full data rate for long time spans is inherently inefficient.

We are exploring solutions to this, which will likely involve returning lower
sampling rates when viewing longer timescales. This will improve efficiency in
queries to the database, improving performance. One likely solution is to use
InfluxDB's ``continuous queries'' mechanism to downsample data and insert it
into another database within InfluxDB. This could be done at various
resolutions, optimized for different length timescales. Dashboards could then
be configured to select the correct database depending on the users needs.
Research remains to be done on this problem and will be ongoing.

\section{Deployment}
\label{sec:deployment}

\ocs{} has been designed as a modular system. Except where Agents must
communicate with specific hardware over dedicated connections or protocols,
Agents can be run anywhere on the network. The use of Docker in the deployment
of an \ocs{} network further facilitates moving Agents to wherever computational
resources are available. Dependencies for a given Agent are bundled within the
pre-built Docker images, reducing the cost of moving the node an Agent is
running on.  This flexibility allows for many possible configurations. In this
section we present a simple, single-node, deployment example, representative of
what might be seen in a test lab. We then discuss the deployment at the
University of California, San Diego (UCSD), which is close to what will be used
on a single SAT. Lastly, we discuss the planned site layout for SO.

\subsection{Simple Deployment Example}

\begin{figure} [ht]
\begin{center}
\begin{tabular}{c}
\includegraphics[width=0.4\linewidth]{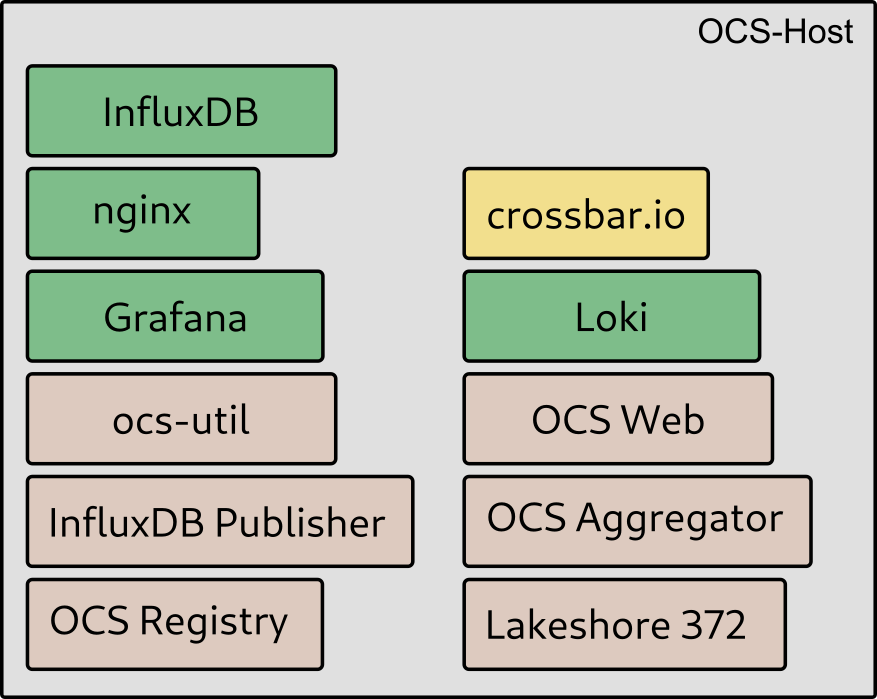}
\end{tabular}
\end{center}
\caption[example]
{ \label{fig:simple-deployment}
A diagram of a simple \ocs{} network. A single node, ``OCS-Host'', runs all
containers, represented by the colored boxes within the node. Connections to
containers within the node are assumed. Users interact via external \ocs{} Clients
(not pictured here), or via OCSWeb and Grafana, served over the NGINX web
server.}
\end{figure}

The simplest \ocs{} deployment is on a single computer running all components
needed for \ocs{}: the crossbar server, any \ocs{} Agents, and supporting monitoring
infrastructure such as Grafana, a web server, and Loki for log aggregation
(described in more detail in Sec. \ref{sec:loki}).  Hardware with dedicated
connections, such as USB connections, connect directly to this machine. Any
networked devices that \ocs{} sends commands to or receives data from must, of
course, be on the same network as this machine. Figure
\ref{fig:simple-deployment} shows a diagram of this simple layout.\footnote{An
example close to this configuration is given in the ``Quickstart'' section of
the \ocs{} documentation.}

A modest desktop machine is capable of running this system. Most systems like
this within SO are equipped with a quad-core CPU from the last 5--7 years and
12--16 GB of RAM. The example shown here could be used for readout and
monitoring of a cryostat in the lab, with temperature control, hardware
permitting. In this setting, data rates are typically low: 10 Hz or less per
Agent.

\subsection{Test Lab Deployment and Development}
\label{sec:test_lab_deployment}

Test institution configurations will often span multiple nodes and operating systems, with network configurations that depend on many factors such as which agents are required to run and network restrictions imposed by the lab or university.
In most cases there is a primary storage node that runs the core set of \ocs{} agents and support containers, including the HK Aggregator and InfluxDB Publisher agents, the registry, the database containers and the Grafana server.
Other nodes that are required to manage specific pieces of hardware such as the Bluefors DR or the SMuRF system are connected on a private subnet.

\begin{figure} [ht]
  \centering
  \includegraphics[width=1.0\textwidth]{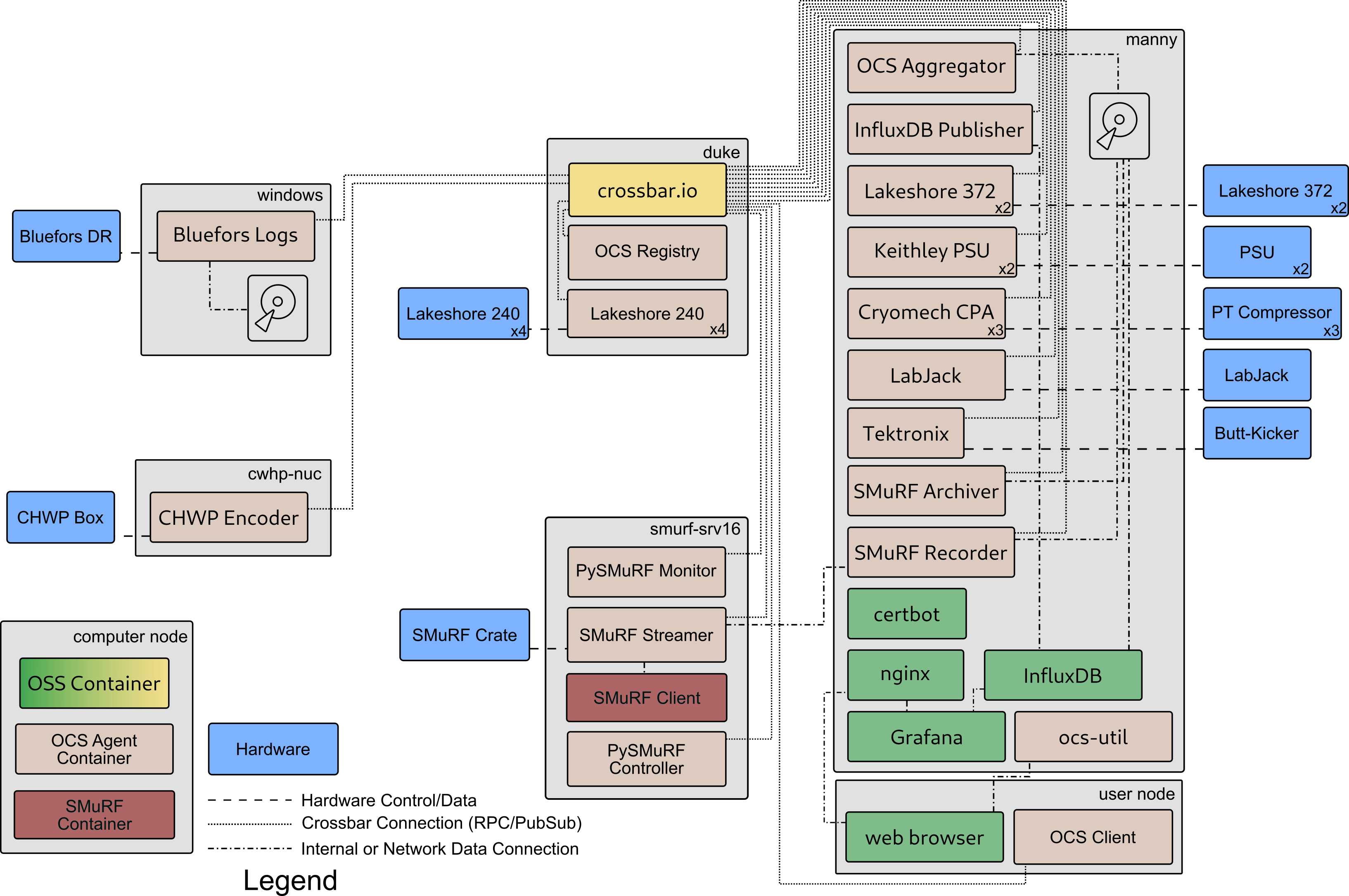}
  \caption{A diagram, similar to Figure \ref{fig:simple-deployment} of the SAT1
development network at UCSD. Connections to the crossbar server are shown, as
well as data connections, and connections to hardware. A legend describing the
various connections and containers is shown in the bottom left. Open source
software (OSS) containers are represented as either green or yellow boxes.}
  \label{fig:ucsd-test-setup}
\end{figure}

The most extensive \ocs{} network in use is located at UCSD to test the integration of the SAT1.
This system is running over 22 agents across five different nodes to control and monitor the SMuRF system, a cryogenic continuously rotating half wave plate (CHWP), Lakeshore devices, power supplies,  the Bluefors DR, and other pieces of hardware. 
The network setup is shown in Figure \ref{fig:ucsd-test-setup}.

\subsection{Planned Site Layout Deployment}

\begin{figure} [ht]
\begin{center}
\begin{tabular}{c}
\includegraphics[width=0.95\textwidth]{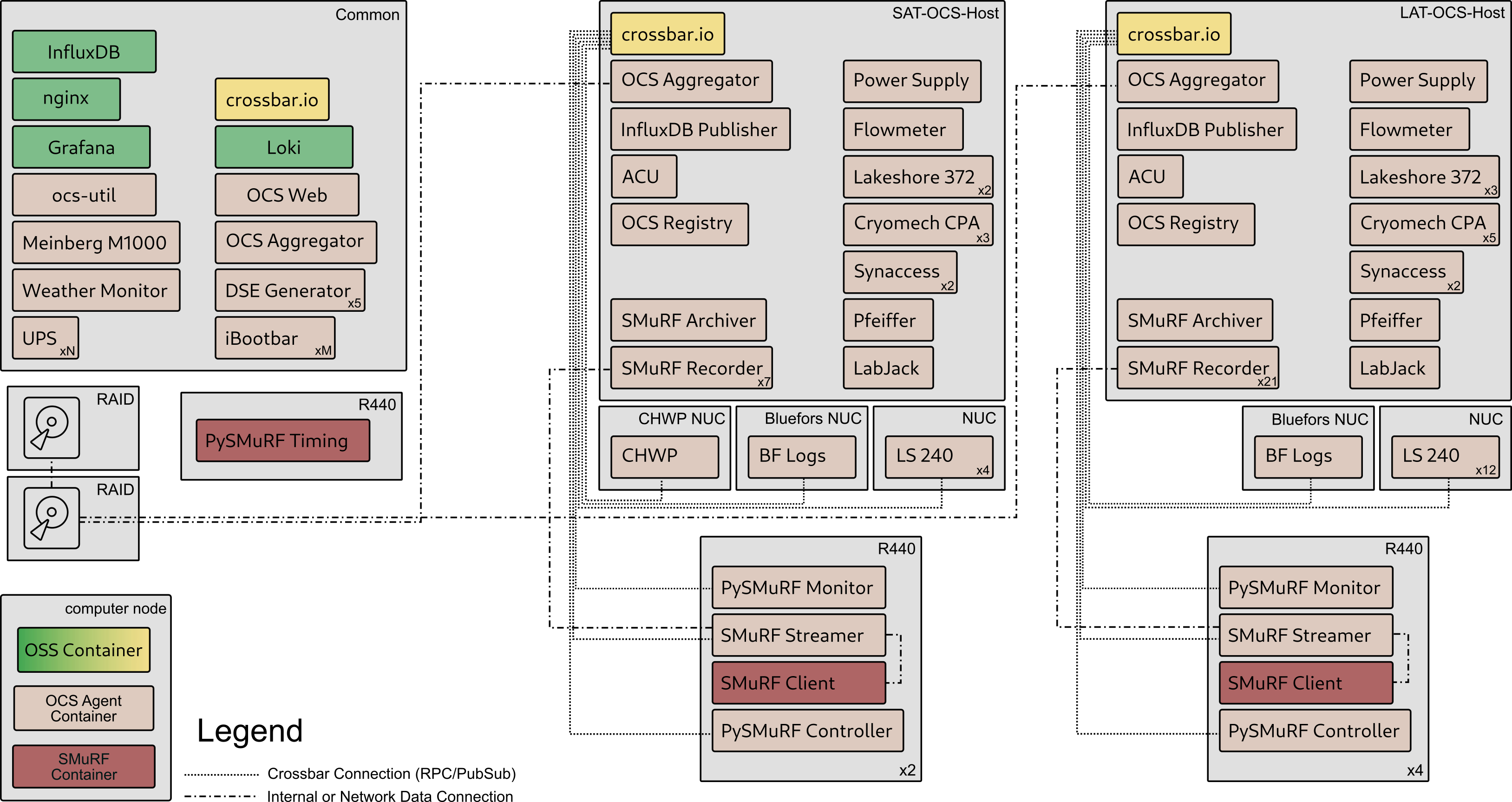}
\end{tabular}
\end{center}
\caption[example]
{\label{fig:site-deployment}
A diagram of the site \ocs{} network. All nodes are connected via the on site
network. Crossbar connections outside of a single node are shown. Connections
within a node are assumed. One instance of an SAT \ocs{} network is shown by the
central group of nodes. At full deployment three of these will be present.
Multiple copies of given Agents are shown with multipliers. The UPS and
iBootbar Agent counts, $N$ and $M$, are still being determined.}
\end{figure}

Once SO is fully deployed, the \ocs{} network on-site will be more than four times
as large as the \ocs{} network discussed in Section \ref{sec:test_lab_deployment}.
Each telescope will have an independent \ocs{} network, meaning each will have its
own crossbar server, HK Aggregator, InfluxDB Publisher, and collection of many
other accompanying Agents. Most of these Agents will run within containers on a
single node (one per telescope). Figure \ref{fig:site-deployment} shows a
diagram of the planned site network.

Each SAT will have three specialized computers on the telescope platform, one
for the cryogenic half-wave plate, one for the Bluefors DR, and one for the
Lakeshore 240s. These will each run agents relevant to the hardware they couple
to. Lastly, there are the SMuRF servers, which run the Agents and SMuRF Client
containers required to read out the detector arrays. There is one of these
servers per SMuRF Crate, meaning there will be two per SAT and four for the
LAT.

The LAT will have very similar Agents to those run on the SAT, though for some
Agents there will be many more running instances of a given Agent, reflecting
the larger overall size of the LAT. The LAT will not include a CHWP node or
associated Agents.

In addition to the four separate \ocs{} networks for each telescope, there will be
one more network for site-wide hardware, such as the Meinberg M1000, used for
site wide GPS synchronized timing, a weather monitor, to monitor the local
weather, and several Agents related to power distribution at the site. These
include UPS and iBootbar Agents, for battery backup of site computing
infrastructure and remote power outlet control, and an Agent for interfacing
with the on-site diesel generators.

Coordination of the \ocs{} Agents distributed across these networks will be
performed by the Observation Sequencer Agent described in Sec
\ref{sec:sequencer} where needed. All output data, whether from the \ocs{}
Aggregator or SMuRF Recorder, will end up on disk in a Redundant Array of
Independent Disks (RAID) hosted on site. Data on this RAID will be available to
local compute nodes which will perform near real time analysis of data for data
quality assurance. Data will be further combined into units suitable for upload
to our data transfer manager (a discussion of which is beyond the scope of this
paper). 

\subsubsection{Site Computing}
The \ocs{} networks described in this section will run on a set powerful,
multi-CPU, high RAM servers located in the control container on site, a
temperature controlled and isolated environment. The SMuRF servers will also
run in this location, and will run on Dell R440 servers. The dedicated NUC
(short for Intel's ``Next Unit of Computing'', a small form factor PC) hardware
on the telescope platforms are ruggedized Intel NUCs equipped with Intel
Celeron processors.

The site will be equipped with a Meinberg M1000 timing system that synchronizes
with GPS satellites and provides precision timing to the site network via the
Precision Time Protocol (PTP) as well as inter-range instrumentation group
(IRIG) timecodes and pulse per second (PPS) timing to the SMuRF master timing
electronics. Each telescope platform will have a Meinberg SyncBox/N2X that
accepts the PTP signal from the network and can be configured to output a
variety of timing signals on three different BNC outputs. These will provide
any precision timing signal required nearby the telescopes, such as for the CHWPs.
The site networking equipment will all be PTP compatible, and appropriately
configured to distribute the PTP timing signal within $\pm 100\,\mathrm{ns}$.

\section{SUMMARY}
\label{sec:summary}

We have presented an overview of the Observatory Control System for the
Simons Observatory. The design and implementation of this distributed control
system is modular and will scale up as we deploy the observatory and beyond.
The use of open source tools, such as \crossbar{}, Grafana, InfluxDB, and Docker,
has allowed us to focus on the core \ocs{} functionality while delivering a
powerful monitoring interface, and convenient deployment options. The
functionality provided by \ocs{} is a critical aspect of the observatory and
will enable the science goals outlined in our forecast
paper.\cite{the_simons_observatory_collaboration_simons_2018}.

\ocs{} is already integrated into the activities of SO through its use in labs
doing hardware development and assembly. These preliminary deployments of \ocs{}
are not only contributing to the ongoing deployment of the experiment but have
also been invaluable for debugging and developing our software. This process
has been assisted by the use of Docker and by the automated testing and
building of Docker images in the CI pipelines. The extensive documentation has
enabled users at these labs to develop their own Agents and extend the
functionality of \ocs{}. \ocs{}, and the SO specific Agents in \socs{}, are both
developed as open source software and are permissively licensed under the BSD
2-Clause license. The software is available on GitHub, and we encourage the
community to try examples detailed within the documentation, to consider using
\ocs{} for their control software, and to contribute to its development.

\appendix

\section{ACRONYMS}
\label{sec:acronyms}

Table \ref{tab:acronyms} shows a list of all acronyms used in this paper.

\begin{table}[h!]
\caption{Acronyms.} 
\label{tab:acronyms}
\begin{center}       
\begin{tabular}{|l|l|}
\hline
\textbf{Acronym} & \textbf{Meaning or Explanation}\\[0pt]
\hline
ACT & The Atacama Cosmology Telescope\\
ACU & Antenna Control Unit\\
API & Application Programming Interface\\
BICEP & Background Imaging of Cosmic Extragalactic Polarization Telescope\\
BNC & Bayonet Neill–Concelman, a common coaxial connector type\\
CI & Continuous Integration\\
CPU & Central processing unit\\
CSS & Cascading Style Sheets\\
CWHP & Cryogenic Half-wave Plate\\
DR & Dilution Refrigerator\\
EPICS & Experimental Physics and Industrial Control System\\
FPGA & Field-programmable gate array\\
\texttt{.g3} & Short term for referring to \texttt{spt3g\_software}\\
GCP & General Control Program, control software used in past CMB experiments\\
GPS & Global Positioning System\\
GUI & Graphical user interface\\
HK & Housekeeping data (i.e. non-detector data)\\
HTML & Hypertext Markup Language\\
HTTP(S) & Hypertext Transfer Protocol (Secure)\\
IRIG & Inter-Range Instrumentation Group\\
LAT & Large Aperture Telescope\\
NUC & Intel's ``Next Unit of Computing"\\
PC & Personal computer\\
PPS & Pulse per second\\
PTP & Precision Time Protocol\\
Pubsub & Publish/Subscribe\\
\ocs{} & Observatory Control System\\
RAID & Redundant Array of Independent Disks\\
RAM & Random-access memory\\
RPC & Remote procedure call\\
SAT & Small Aperture Telescope\\
SCF & OCS Site Configuration File\\
SLAC & SLAC National Accelerator Laboratory\\
SMuRF & SLAC Microresonator Radio Frequency\\
SPT & South Pole Telescope\\
SO & Simons Observatory\\
\socs{} & Simons Observatory Control System\\
SPT3G & The South Pole Telescope 3rd Generation, can also refer to the spt3g software package\\
TCP & Transmission Control Protocol\\
TES & Transition Edge Sensor\\
UCSD & University of California, San Diego\\
UDP & User Datagram Protocol\\
$\mu\mathrm{MUX}$ & Microwave multiplexing\\
UPS & Uninterruptible power supply\\
USB & Universal Serial Bus\\
WAMP & Web Application Messaging Protocol\\
YAML & Yet Another Markup Language\\
\hline
\end{tabular}
\end{center}
\end{table}

\acknowledgments
This work was funded by the Simons Foundation (Award \#457687). ZX is supported
by the Gordon and Betty Moore Foundation. We would like to thank the
communities behind the multiple open source packages in use with \ocs{}.
 
\bibliography{report,main}
\bibliographystyle{spiebib}

\end{document}